\documentclass[11pt]{article}
\usepackage{fullpage}
\usepackage{rpmacros}
\usepackage[T1]{fontenc}
\usepackage{gitinfo}
\usepackage[usenames,dvipsnames]{xcolor}
\usepackage{stmaryrd}
\usepackage{xfrac}
\usepackage{mathpazo}
\usepackage{bm}
\usepackage{todonotes}
\usepackage{lipsum}
\usepackage{setspace}
\usepackage{scrextend}
\usepackage{dsfont}
\usepackage[shortlabels]{enumitem}

\newif\ifdraft
\draftfalse

\RequirePackage[colorlinks=true]{hyperref}
\hypersetup{
	linkcolor=[rgb]{0.3,0.3,0.6},
	citecolor=[rgb]{0.2, 0.6, 0.2},
	urlcolor=[rgb]{0.6, 0.2, 0.2}
}
\usepackage{amsthm}
\usepackage{thmtools,thm-restate}

\numberwithin{equation}{section}
\declaretheoremstyle[bodyfont=\it,qed=\qedsymbol]{noproofstyle}

\declaretheorem[numberlike=equation]{observation}

\declaretheorem[name=Observation,numbered=no]{observation*}

\declaretheorem[numberlike=equation]{theorem}

\declaretheorem[name=Theorem,numbered=no]{theorem*}

\declaretheorem[numberlike=equation]{lemma}
\declaretheorem[name=Lemma,numbered=no]{lemma*}

\declaretheorem[numberlike=equation]{corollary}
\declaretheorem[name=Corollary,numbered=no]{corollary*}

\declaretheorem[name=Proposition,numbered=no]{proposition*}

\declaretheorem[numberlike=equation]{claim}
\declaretheorem[name=Claim,numbered=no]{claim*}

\declaretheorem[name=Conjecture,numbered=no]{conjecture*}

\declaretheorem[name=Question,numbered=no]{question*}

\declaretheoremstyle[bodyfont=\it,qed=$\lozenge$]{defstyle} 

\declaretheorem[numberlike=equation,style=defstyle]{definition}
\declaretheorem[unnumbered,name=Definition,style=defstyle]{definition*}

\declaretheorem[unnumbered,name=Example,style=defstyle]{example*}

\declaretheorem[unnumbered,name=Notation=defstyle]{notation*}

\declaretheorem[unnumbered,name=Construction,style=defstyle]{construction*}

\declaretheorem[unnumbered,name=Remark,style=defstyle]{remark*}

\newcommand{\prob}[2]{\mathop{\mathrm{Pr}}_{#1}\insquare{#2}}

\newcommand{\WtDist}{\operatorname{WtDist}}

\newcommand{\wt}{\operatorname{wt}}

\newcommand{\RM}{\operatorname{RM}}
\newcommand{\Span}{\operatorname{span}}

\newcommand{\BEC}{\operatorname{BEC}}
\newcommand{\BSC}{\operatorname{BSC}}

\newcommand{\I}{\mathbb{I}} % For the vanishing ideal
\newcommand{\closure}{\operatorname{Closure}}

\usepackage{nth}
\usepackage{intcalc}
\usepackage{etoolbox}
\usepackage{xstring}

\usepackage{ifpdf}
\ifpdf
\else
\usepackage[quadpoints=false]{hypdvips}
\fi

\newcommand{\ECCCurl}[2]{http://eccc.hpi-web.de/report/\ifnumcomp{#1}{>}{93}{19}{20}#1/#2/}

\newcommand{\shortECCC}[2]{\texttt{\href{http://eccc.hpi-web.de/report/\ifnumcomp{#1}{>}{93}{19}{20}#1/#2/}{eccc:TR#1-#2}}}

\newcommand{\parseECCC}[1]{% Takes a string of the form TRxx/xxx or
\StrSubstitute{#1}{TR}{}[\tmpstring]%
\IfSubStr{\tmpstring}{/}{ %assuming string is of the form TRxx/xxx
\StrBefore{\tmpstring}{/}[\ecccyear]%
\StrBehind{\tmpstring}{/}[\ecccreport]%
}{% assuming string is of the form TRxx-xxx
\StrBefore{\tmpstring}{-}[\ecccyear]%
\StrBehind{\tmpstring}{-}[\ecccreport]%
}%
\shortECCC{\ecccyear}{\ecccreport}}

\ifdraft
\newcommand{\RPnote}[1]{\textcolor{BrickRed}{\guillemotleft RP: #1 \guillemotright}}
\newcommand{\PHnote}[1]{\textcolor{OliveGreen}{\guillemotleft PH: #1 \guillemotright}}
\newcommand{\SBnote}[1]{\textcolor{NavyBlue}{\guillemotleft SB: #1 \guillemotright}}
\newcommand{\SSnote}[1]{\textcolor{Purple}{\guillemotleft SS: #1 \guillemotright}}
\else
\newcommand{\RPnote}[1]{}
\newcommand{\PHnote}[1]{}
\newcommand{\SBnote}[1]{}
\newcommand{\SSnote}[1]{}
\fi

\renewcommand{\epsilon}{\varepsilon}
\renewcommand{\phi}{\varphi}

\newcommand{\ignore}[1]{}
\newcommand{\ehref}[1]{\href{mailto:#1}{#1}}

\onehalfspace
\allowdisplaybreaks

\makeatletter
 \def\and{%
   \end{tabular}%
   \hskip 1em \@plus.17fil\relax
   \begin{tabular}[t]{c}}
\makeatother

\title{Vanishing Spaces of Random Sets\\
and Applications to Reed-Muller Codes}
\author{
    {Siddharth Bhandari\thanks{Simons Institute for the Theory of Computing, Berkeley, USA, \ehref{siddharth.bhandari@berkeley.edu}. Research supported by the Simons-Berkeley Postdoctoral Fellowship.}}
    \and 
    {Prahladh Harsha\thanks{Tata Institute of Fundamental Research, Mumbai, India, \ehref{prahladh@tifr.res.in}. Research done when the author was visiting the Simons Institute for the Theory of Computing. Research supported in part by the Department of Atomic
    Energy, Government of India, under project 12-R\&D-TFR-5.01-0500 and the Swarnajayanti Fellowship.}}
    \and 
    {Ramprasad Saptharishi\thanks{Tata Institute of Fundamental Research, Mumbai, India, \ehref{ramprasad@tifr.res.in}. Research supported by the Department of Atomic Energy, Government of India, under project 12-R\&D-TFR-5.01-0500 and the Ramanujan Fellowship of the DST.}}
    \and 
    {Srikanth Srinivasan\thanks{Aarhus University, Aarhus, Denmark, \ehref{srikanth@cs.au.dk}. Supported by Startup grant from Aarhus University.}}
}
\date{ }
\begin{document}
\maketitle

\begin{abstract}
    \noindent
    We study the following natural question on random sets of points in $\F_2^m$:
    \begin{quote}
        Given a random set of $k$ points
        $Z=\{z_1, z_2, \dots, z_k\} \subseteq \F_2^m$, what is the dimension
        of the space of degree at most $r$ multilinear polynomials that vanish
        on all points in $Z$?
    \end{quote}
    We show that, for $r \leq \gamma m$ (where $\gamma > 0$ is a small, absolute constant) and $k = (1-\epsilon) \cdot \binom{m}{\leq r}$ for any constant $\epsilon > 0$, the space of degree at most $r$ multilinear polynomials vanishing on a random set $Z = \set{z_1,\ldots, z_k}$ has dimension exactly $\binom{m}{\leq r} - k$ with probability $1 - o(1)$. This bound shows that random sets have a much smaller space of degree at most $r$ multilinear polynomials vanishing on them, compared to the worst-case bound (due to Wei (IEEE Trans. Inform. Theory, 1991)) of $\binom{m}{\leq r} - \binom{\log_2 k}{\leq r} \gg \binom{m}{\leq r} - k$. 

    Using this bound, we show that high-degree Reed-Muller codes ($\RM(m,d)$ with $d > (1-\gamma) m$) ``achieve capacity'' under the Binary Erasure Channel in the sense that, for any $\epsilon > 0$, we can recover from $(1 - \epsilon) \cdot \binom{m}{\leq m-d-1}$ random erasures with probability $1 - o(1)$. This also implies that $\RM(m,d)$ is also efficiently decodable from $\approx \binom{m}{\leq m-(d/2)}$ random errors for the same range of parameters. 
\end{abstract}

\section{Introduction}

The Reed-Muller (RM) code is one of the most basic error-correcting codes
studied in coding theory, first introduced by Muller~\cite{Muller1954}
and Reed~\cite{Reed1954} in 1954. Stated in the language of
polynomials, the RM code with parameters $m$ and $r$ for
positive integers $m > r$, denoted by $\RM(m,r)$, is the code whose
codewords are evaluations of $m$-variate multilinear polynomials of
degree at most $r$ over the vector space $\F_2^m$. Despite these being
one of the earliest codes discovered in coding theory, several
properties of these codes (weight-distribution, capacity-achieving on
any binary memory-less symmetric (BMS) channel) are not yet
fully-understood.

A natural problem that arises while investigating the Shannon-capacity
of RM code is the following:

\begin{quote}
\emph{Given a set $Z$ of $k$ points
$Z=\{z_1, z_2, \dots, z_k\} \subseteq \F_2^m$, what is the dimension
of the space $\I_r(Z)$ of degree at most $r$ multilinear polynomials that vanish
on all points in $Z$?}
\end{quote}

To understand the connection to RM codes, we give an equivalent
description of this problem in terms of the parity check matrix of the
RM code. Let $E(m,r)$ be the $\binom{m}{\leq r} \times 2^m$-matrix
whose columns are indexed by elements in $\F_2^m$ and rows by
$m$-variate multilinear monomials of degree at most $r$. The $z^{th}$
column of $E(m,r)$ for $z \in \F_2^m$ is the column vector $z^{(r)}$
consisting of evaluations of all multilinear $m$-variate degree
$\leq r$ monomials at the point $z$. It is not hard to see that
$E(m,r)$ is the parity check matrix of the Reed-Muller code
$\RM(m,m-r-1)$. An equivalent formulation of the above question in
terms of the matrix $E(m,r)$ is the following: given a set $Z$ of $k$
points $Z=\{z_1, z_2, \dots, z_k\} \subseteq \F^m_2$, what is the rank
of the $\binom{m}{\leq r} \times |Z|$ sub-matrix $E(m,r)_Z$ of
$E(m,r)$ obtained by picking the columns corresponding to the points
in $Z$. 

While studying the generalized Hamming weights of RM codes over $\F_2$,
Wei~\cite{Wei1991} proved worst-case bounds for the above problem. In
its simplest form, these worst-case bounds show that if $|Z| = 2^\ell$
for some $0 \leq \ell \leq m$, then
$\dim(\I_r(Z)) \leq \binom{m}{\leq r} - \binom{\ell}{\leq r}$. 
These worst-case bounds were then generalized to arbitrary
fields by Keevash and Sudakov~\cite{KeevashS2005} and have had applications in combinatorics as well as  theoretical computer science. Ben-Eliezer, Hod and Lovett~\cite{BenEliezerHL2012}
reproved the above form of this bound and used it to study the correlation
of random polynomials with lower-degree polynomials. Nie and
Wang~\cite{NieW2015} used these bounds bound to prove extensions of the
Kakeya Theorem. Chen, De, and Vijayaraghavan~\cite{ChenDV2021} used these bounds to analyze their algorithms for learning mixtures of subspaces over finite fields. In the context of RM codes, Abbe, Shpilka and
Wigderson~\cite{AbbeSW2015} used these worst-case bounds to show that RM codes of high
rate  ($r = m - o(\sqrt{m/\log m})$) achieve Shannon-capacity over the
BEC channel. 

These bounds due to Wei only prove extremal (i.e., worst-case) bounds
on $\dim(\I_r(Z))$. The tightness of this bound is witnessed when $Z$
is a subcube of size $2^\ell$, where
$\dim(\I_r(Z) = \binom{m}{\leq r} - \binom{\ell}{\leq r}$ which is
considerably larger than
$\binom{m}{\leq r} - |Z| = \binom{m}{\leq r} - 2^\ell$.  We note that
$\binom{m}{\leq r} - |Z|$ is a lower bound on the dimension of
$\I_r(Z)$ since each point in $Z$ can reduce the dimension by at most
$1$. It is thus natural to ask how large is $\dim(\I_r(Z))$ for sets $Z$
other than a subcube. In particular, how does $\dim(\I_r(Z))$ compare
to $\binom{m}{\leq r} - |Z|$ when $Z$ is a random set of points of
size $K < \binom{m}{\leq r}$.  Our main result shows that for small $r$ (more precisely, for
$r \leq \gamma m$ for a small, but absolute, constant $\gamma > 0$)
and $|Z| = (1-\epsilon) \cdot \binom{m}{\leq r}$ for any constant
$\epsilon \in (0,1)$, a random set $Z$ behaves very differently from a
subcube and the dimension of $\I_r(Z)$ is as small as it can be, namely
$\binom{m}{\leq r} - |Z|$.

\begin{restatable}[dimension of degree-$r$ vanishing space of random set]{theorem}{RestateVanishingIdealRandomSet}
    \label{thm:vanishing-space-random-set}
    There exists a constant $\gamma_0>0$ such that for all $\epsilon>0$ the following is true. Let $K = (1 - \epsilon) \cdot \binom{m}{\leq r}$ and $r  < \gamma_0 m$. Then, 
    \[
        \prob{Z \subseteq \F_2^m}{\dim(\I_r(Z)) =
          \binom{m}{\leq r} -  K} = 1-o(1)
    \]
    where $Z = \set{z_1,\ldots, z_K}$ is a uniformly random set of $K$ distinct points in $\F_2^m$. In other words, the probability that the set of columns $\{z_1^{(r)},z_2^{(r)},\ldots, z_K^{(r)}\}$ is not linearly independent is $o(1)$, where $z^{(r)}$ denotes the vector consisting of evaluations of all multilinear $m$-variate degree $\leq r$ monomials at the point $z$.
\end{restatable}

 Abbe, Shpilka and Wigderson~\cite{AbbeSW2015} also proved a similar result for a smaller range of $r$, namely $r = o(\sqrt{m/\log m})$ using Wei's extremal bound. We note that if $\gamma_0$ is chosen sufficiently small such that $K \leq \binom{m}{\leq r} = o(2^{m/2})$, then it does not matter if the points $z_1,\ldots,z_k$ are chosen with or without replacement since the two distributions are then $o(1)$-indistinguishable. When the points are chosen with replacement, then one cannot expect to improve the range of $r$ for which the above theorem holds (up to the choice of the constant $\gamma_0$) since if $K \gg 2^{m/2}$, then with high probability two of the $z_i$'s are equal in which case $\I_r(Z)$ will definitely have dimension strictly larger than $\binom{m}{\leq r} -  K$.

As one would expect, this average-case bound on $\dim(\I_r(Z))$ leads
to a better understanding of the Shannon-capacity of RM codes over the
high-rate regime. 

\paragraph{Shannon-capacity of Reed-Muller codes} 
Reed-Muller codes were shown to achieve Shannon-capacity over the BEC
channel in the extremal regimes (low rate $r =
o(m)$ and high rate $r = m - o(m)$) by Abbe, Shpilka and Wigderson~\cite{AbbeSW2015} and
in the constant rate regime ($r = m/2 \pm O(\sqrt{m})$) by Kudekar et
al~\cite{KudekarKMPSU2017}. More recently, Reeves
and Pfister~\cite{ReevesP2021} showed that RM codes over the constant rate regime ($r =
m/2 \pm O(\sqrt{m})$) achieve Shannon-capacity over any BMS
channel. Despite all this progress, the general question of whether RM
codes achieve capacity over the entire regime ($1 \leq r < m-1$) and
over any BMS channel remains open.

The results of Kudekar et al~\cite{KudekarKMPSU2017} and Reeves and
Pfister~\cite{ReevesP2021} are obtained using Boolean function
analysis while the results of Abbe, Shpilka and Wigderson for the
Shannon-capacity over the BEC channel of RM codes in the low-rate
regime $(r=o(m))$ were obtained using the weight-distribution bound of
Kaufman, Lovett and Porat~\cite{KaufmanLP2012}. Subsequent
improvements in the weight distribution due to Sberlo and Shpilka~\cite{SberloS2020} led
to the Shannon-capacity of RM codes over the BEC channel for a wider range of the degree
parameter $r$, more
precisely for $r = \gamma m$ for $\gamma < 1/70$.

The Shannon-capacity of RM over the high-rate regime
$(r = m - o(\sqrt{m/\log m})$ were also proved by Abbe, Shpilka and
Wigderson using Wei's worst-case bounds on $\dim(\I_r(Z))$. In
particular, these latter results do not use the weight distribution of
RM codes and hence the subsequent improvements in our understanding of
the weight distribution of RM codes due to
Samorodnitsky~\cite{Samorodnitsky2020} and
Sberlo-Shpilka~\cite{SberloS2020} did not lead to an improved
understanding of the Shannon-capacity of RM codes in the high-rate
regime. We rectify this gap by giving an alternate bound for the
Shannon-capacity in terms of the our average-case bound on
$\dim(\I_r(Z))$ and widen the range of the degree parameter $r$ for
which high-rate RM codes achieve Shannon-capacity over the BEC
channel.  

\begin{restatable}[high-degree RM codes under erasures]{theorem}{RestateHighDegRMBEC}
    \label{thm:high-deg-RM-under-BEC}
    There exists a constant $\gamma_0>0$ such that for all $\epsilon>0$ the following is true. Let $m,d$ be growing parameters with $d > m (1 - \gamma_0)$. Then, the code $\RM(m,d)$ can correct $K = (1-\epsilon) \binom{m}{\leq m-d-1}$ random errors with probability $1-o(1)$. 
\end{restatable}

Abbe, Shpilka and Wigderson~\cite{AbbeSW2015} and
Saptharishi, Shpilka and Volk~\cite{SaptharishiSV2017} showed how to
reduce  the resilience of certain RM codes under the BSC to
the resilience of appropriate RM codes under the BEC. Using
this reduction, we obtain the following corollary of the above
theorem for the BSC channel. 

\begin{restatable}[high-degree RM codes under errors]{corollary}{RestateHighDegRMBSC}
    \label{cor:high-deg-RM-under-BSC}
    There exists a constant $\gamma_0>0$ such that for all $\epsilon>0$ the following is true. Let $m,r$ be growing parameters with $r < \gamma_0 m$. Then, the code $\RM(m,m-2r-2)$ can be efficiently decode from $K = (1 - \epsilon)\binom{m}{\leq r}$ random errors.
\end{restatable}
See \autoref{sec:shannon} for further discussion on Shannon-capacity and what it means for high-rate RM codes to achieve it over the BEC and BSC channels. 

We remark that our proof methods work for any linear code, not necessarily the RM code, provided one has good bounds on the weight distribution of the dual code. 

\subsection{Proof Overview}

Recall that \autoref{thm:vanishing-space-random-set} is a statement about the dimension of the degree-$r$ vanishing space of a uniformly random subset $Z$ of $\F_2^m$ of size $K$ where $K=(1-\varepsilon)\cdot \binom{m}{\leq r}.$ In the regime of parameters we are interested in (i.e., $r< \gamma_0 m$ for a small enough constant $\gamma_0$), this is more or less equivalent to\footnote{In the sense that the two distributions have small statistical distance.} choosing $K$ points $z_1,\ldots,z_K$ independently and uniformly from $\F_2^m.$ We assume that $Z$ is defined this way for the rest of this section.

To argue about the dimension of the degree-$r$ vanishing space $\I_r(Z)$, we instead argue about the \emph{size} $S$ of $\I_r(Z)$. Note that since $\dim(\I_r(Z))\geq D := \binom{m}{\leq r} - K$, this set has size \emph{at least} $2^D$, and has size at least $2^{D+1}$ if $\dim(\I_r(Z)) > D.$ In light of this, it is sufficient to show that 
\[\E[S] = \exp_2(D)(1+o(1)).\]

Estimating $\E[S]$ turns out to be very closely related to a recent result of Sberlo and Shpilka~\cite{SberloS2020}, who prove strong results on the parameters of capacity-achieving Reed-Muller codes in the low-degree setting. 

More precisely, we can easily see that the probability that a uniformly random polynomial $P\in \RM(m,r)$ belongs to $\I_r(Z)$ is exactly $(1-\wt(P))^K$ where $\wt(P)$ is the \emph{fractional Hamming weight of $P$} (i.e., the fraction of points where it does not vanish). Thus, we have
\[
\E[S] = \sum_{P\in \RM(m,r)} (1-\wt(P))^K.
\]

Now, while there are polynomials $P\in \RM(m,r)$ of very small weight\footnote{It is a standard fact (e.g. by the Schwartz-Zippel lemma) that the minimum weight of a non-zero polynomial from $\RM(m,r)$ is $2^{-r}$.}, \emph{most} polynomials in $\RM(m,r)$ have weight close to $1/2,$ which should indicate that the sum is close to $2^{-K}$ as required. However, a careful analysis is required, as the contribution of a polynomial $P$ increases exponentially as its weight decreases.

It turns out that Sberlo and Shpilka~\cite{SberloS2020} analyzed a very similar quantity in their recent work. More precisely they showed that for any constant $\delta > 0$ (and also a large range of sub-constant $\delta$), we have
\[
\sum_{P\in \RM(m,r)\setminus \set{\mathbf{0}}} (1-\wt(P))^{(1+\delta)\cdot \binom{m}{\leq r}} = o(1).
\]
While this is very closely related to our result, we are not able to recover the exact bound we need using this inequality. 

However, the technical lemmas used to derive the above result are strong results on the \emph{weight distribution} of $\RM(m,r)$, which are upper bounds on the number of polynomials of small weight. Using these upper bounds and carrying out the relevant computations yields \autoref{thm:vanishing-space-random-set}.

The application to resilience under the BEC and BSC (\autoref{thm:high-deg-RM-under-BEC} and \autoref{cor:high-deg-RM-under-BSC}) follow from \autoref{thm:vanishing-space-random-set} in a straight-forward manner from the works of Abbe, Shpilka and Wigderson~\cite{AbbeSW2015} and Saptharishi, Shpilka and Volk~\cite{SaptharishiSV2017}. 

\subsection*{Organisation}

We discuss some notation and preliminaries in \autoref{sec:notation} and proceed to the proof of \autoref{thm:vanishing-space-random-set} in \autoref{sec:proof_of_main_thm}. We then present the applications to proving the resilience of RM codes under BEC and BSC in \autoref{sec:resilience_to_channels}. 

\section{Notation and preliminaries}
\label{sec:notation}

\begin{enumerate}
    \item We use $[m]$ to represent the set $\{1,2,3,\ldots,m\}$ and the expression $\binom{m}{\leq r}$ to denote the sum 
    \[
        \binom{m}{0} + \cdots + \binom{m}{r}.
    \]
    \item  We denote by $\binom{[m]}{ r}$ and $\binom{[m]}{\leq r}$ the sets $\{S \subseteq [m] \colon |S| = r\}$ and  $\{S\subseteq[m]\colon |S|\leq r \}$ respectively. 
    \item We shall abuse notation and use $\RM(m,r)$ to denote the vector space of all $m$-variate multivariate polynomials in $\F_2[x_1,\ldots, x_m]$ of degree at most $r$. Throughout the paper, the parameter $m$ will be unchanged and we will avoid mentioning it for brevity. Let $n = 2^m$.
    \item For parameters $m,r$, define the matrix $E(m,r)$ as the $\binom{m}{\leq r} \times 2^m$-matrix whose columns are indexed by elements in $\F_2^m$ and rows by multlinear $m$-variate monomials of degree at most $r$ and whose $z$-th column is the vector consisting of the \emph{evaluation} of all multilinear $m$-variate degree $\leq r$ monomials on $z$.
    \item For a polynomial $P\in \RM(m,r)$, we use $\wt(P)$ to denote the \emph{fractional Hamming weight}:
    \[
        \wt(P) := \frac{\abs{ \setdef{z\in\F_2^m}{P(z) \neq 0} }}{2^m}
    \]
    We shall say that a polynomial $P \in \RM(m,r)$ is $\eta$-biased if $\abs{\wt(P) - \sfrac{1}{2}} \geq \eta$. We shall also use $\RM_\eta(m,r)$ to denote the set of all $\eta$-biased polynomials in $\RM(m,r)$. 
    \item For a real number $\alpha \in (0,1)$, we denote by $\WtDist_{m,r}(\alpha)$ the number of polynomials of (fractional) weight at most $\alpha$ in $\RM(m,r)$, i.e., $\WtDist_{m,r}(\alpha) := |\{P \in \RM(m,r) \colon \wt(p) \leq \alpha \}|.$
    \item All complexity notations used in the paper are with respect to $m$ as the growing parameter.
\end{enumerate}

\subsection{Weight distribution bounds}

We use the following bounds on the weight-distribution of Reed-Muller codes due to Sberlo and Shpilka~\cite{SberloS2020}. 

\begin{theorem}[Sberlo and Shpilka~\cite{SberloS2020}: bounds for low-weight codewords]
    \label{thm:SS-low-wt}
    For any $\ell \geq 1,$ we have 
    \[
    \WtDist_{m,r}(2^{-\ell}) \leq \exp_2\left(O(m^4) + 17\cdot (c_\gamma \ell + d_\gamma)\cdot \gamma^{\ell-1}\cdot \binom{m}{\leq r}\right)
    \]
    where $c_\gamma = 1/(1-\gamma)$ and $d_{\gamma} = \frac{(2-\gamma)}{(1-\gamma)^2}.$
    
    In particular, if $\gamma\leq (1/2)$, we have
    \begin{equation}
    \label{eq:SS-low-wt}
    \WtDist_{m,r}(2^{-\ell}) \leq \exp_2\left(O(m^4) + O(\ell\cdot \gamma^{\ell-1})\cdot  \binom{m}{\leq r}\right).
    \end{equation}
\end{theorem}
    
\begin{theorem}[Sberlo and Shpilka~\cite{SberloS2020}: bounds for medium-weight codewords]
    \label{thm:SS-high-wt}
    Assume that $\gamma \in (0,(1/2)-\Omega(1)).$ For a positive integer $\ell$ such that $\ell/m$ upper bounded by a small enough constant\footnote{\cite[Theorem 1.3]{SberloS2020} only guarantees this for $\ell = o(m)$ but \cite[Remark 1.1]{SberloS2020} after the theorem statement says that in this setting it holds for $\ell = \Omega(m).$}, we have
    \begin{equation}
    \label{eq:SS-high-wt}
    \WtDist_{m,r}\left(\frac{1}{2} - 2^{-\ell}\right) \leq \exp_2\left(O(m^4) + (1-2^{-c(\gamma,\ell)})\cdot \binom{m}{\leq r}\right)
    \end{equation}
    where $c(\gamma,\ell) = O(\max\{\gamma^2 \ell, \gamma\}).$
\end{theorem}

\section{Degree-\texorpdfstring{$\boldsymbol{r}$}{r} vanishing spaces and closures}
\label{sec:proof_of_main_thm}

We first define the notion of a \emph{vanishing space}. This is similar to the notion of \emph{vanishing ideals} in basic algebraic geometry but we refer to them as \emph{vanishing space} instead to stress that we are studying them as a vector space and not an ideal. 

\begin{definition}[degree-$r$ vanishing spaces]
    \label{defn:deg-r-vanishing-spaces}
    For a set $Z \subseteq \F_2^m$, we use $\I_r(Z)$ to denote the \emph{degree-$r$ vanishing space} defined as
    \[
        \I_r(Z) := \setdef{P\in \RM(m,r)}{P(z) = 0\;\text{for all $z\in Z$}}.\qedhere
    \]
\end{definition}

Related to the vanishing ideals is also the notion of the \emph{degree-$r$ closure} (similar to the Zariski closure in standard algebraic geometry but restricted to the setting of $\F_2^m$), which is the set of points on which every polynomial in the degree-$r$ vanishing space vanishes. 

\begin{definition}[degree-$r$ closure]
    \label{defn:deg-r-closure}
    For a set $Z \subseteq \F_2^m$, we use $\closure_r(Z)$ to denote the \emph{degree-$r$ closure} of $Z$ defined as
    \[
        \closure_r(Z) := \setdef{u\in \F_2^m}{P(u) = 0 \text{ for all $P\in \I_r(Z)$}}.\qedhere
    \]
\end{definition}
The above notion can be equivalently defined as the set of all $u \in \F_2^m$ such that column of $E(m,r)$ indexed by $u$ is in the span of columns of $E(m,r)$ indexed by $z\in Z$. That is,
\[
    \closure_r(Z) = \setdef{u\in \F_2^m}{u^{(r)} \in \Span\setdef{z^{(r)}}{z\in Z}},
\]
where $z^{(r)}$ denotes the $\binom{m}{\leq r}$-dimension vector of evaluations of all $m$-variate degree at most $r$ monomials at the point $z$. 

\medskip

For any set $Z \subseteq \F_2^m$ of size $k$, note that $\I_r(Z)$ is a vector space of dimension at least $\binom{m}{\leq r} - k$, as each constraint $P(z) = 0$ adds one homogeneous linear constraint on the ambient space $\RM(m,r)$. In fact, it is easy to see that $\I_r(Z)$ has rank $\binom{m}{\leq r} - k$ if and only if each point of $Z$ is not in the closure of the previous points. 

\begin{observation}[vanishing ideals of minimal rank]
    Let $Z = \set{z_1,\ldots, z_k} \subseteq \F_2^m$. Then, 
    \[
        \dim \I_r(Z) = \binom{m}{\leq r} - k \Longleftrightarrow \inparen{\forall i = 1,\ldots, k-1\;\colon\; z_{i} \notin \closure_r(\set{z_1,\ldots, z_{i-1}})}. \hfill\qedsymbol
    \]
\end{observation}

\subsection{Dimension of the degree-\texorpdfstring{$\boldsymbol{r}$}{r} vanishing spaces of random sets}

In this section, we will be interested in studying the vanishing spaces and closures of a random set $Z$ obtained by picking $K = (1 - \epsilon)\binom{m}{\leq r}$ points from $\F_2^m$ (where $\epsilon > 0$ is some constant). We restate \autoref{thm:vanishing-space-random-set} below.

\RestateVanishingIdealRandomSet*

\bigskip

\noindent
We will use the following lemma to prove the above theorem.

\begin{lemma}[size of degree $r$-vanishing space of random set]
\label{lem:dim_to_size_of_vanishing_ideal}
   There exists a constant $\gamma_0>0$ such that for all $\epsilon>0$ the following is true. Let $K = (1 - \epsilon) \cdot \binom{m}{\leq r}$ and $r  < \gamma_0 m$. Then, 
    \[
        \E \left[\frac{|\I_r(Z)|}{\exp_2\left(\binom{m}{\leq r}\right)}\right]= 2^{-K}(1+o(1))
    \]
    where $Z = \set{z_1,\ldots, z_K}$ is a uniformly random set of $K$ distinct points in $\F_2^m$. 
\end{lemma}

We first use the above lemma to prove \autoref{thm:vanishing-space-random-set}. \autoref{lem:dim_to_size_of_vanishing_ideal} is proved in \autoref{sec:main-lemma}.

\begin{proof}[Proof of \autoref{thm:vanishing-space-random-set}]
    Let $p$ be the probability that $\dim(\I_r(Z))=\binom{m}{\leq r}-K$. Note that $\dim(\I_r(Z))$ is always \emph{at least} $\binom{m}{\leq r} - K$, as the space $\I_r(Z)$ is a subspace of $RM(m,r)$ defined by $K$ linear equations. Thus, 
    \begin{align*}
         \E \left[\frac{|\I_r(Z)|}{\exp_2\left(\binom{m}{\leq r}\right)}\right] &= \E\left[\exp_2\left(\dim(\I_r(Z))-\binom{m}{\leq r}\right)\right] \\
         &\geq p\cdot \exp_2(-K) + (1-p)\cdot 2\cdot \exp_2(-K).
    \end{align*}
    Combining this with \autoref{lem:dim_to_size_of_vanishing_ideal} we get that 
    \[
        p\cdot \exp_2(-K) + (1-p)\cdot 2\cdot \exp_2(-K) \leq 2^{-K}(1+o(1)) 
    \]
    which gives $1 - p \leq o(1)$. Hence, we get that $p = 1-o(1)$. \qedhere
\end{proof}

\subsection{Proof of Lemma~\ref{lem:dim_to_size_of_vanishing_ideal}}
\label{sec:main-lemma}

We start with reformulating the problem of bounding size of $\I_r(Z)$ to computing a certain weighted sum of the polynomials in $\RM(m,r).$

Let $Q$ be a uniformly random polynomial chosen from $\RM(m,r)$. Then,
    \begin{align*}
    \E \left[\frac{|\I_r(Z)|}{\exp_2\left(\binom{m}{\leq r}\right)}\right] &= \E_{Z,Q}\left[ \mathds{1}[Q \in \I_r(Z)] \right]\\
    &=\E_{Q}\left[\E_{Z}[ \mathds{1}[Q \in \I_r(Z)] ]\right]\\
    &=\sum_{P\in \RM(m,r)} \exp_2\left(-\binom{m}{\leq r}\right)\cdot \frac{\binom{(1-\wt(P))2^m}{K}}{\binom{2^m}{K}}\\
    &=\exp_2(-K)\cdot \left[ \sum_{P\in \RM(m,r)} \exp_2\left(-\epsilon\cdot \binom{m}{\leq r}\right)\cdot \frac{\binom{(1-\wt(P))2^m}{K}}{\binom{2^m}{K}}  \right].\qedhere
\end{align*}

\noindent
To complete the proof of \autoref{lem:dim_to_size_of_vanishing_ideal} we therefore need the following claim.

\begin{claim} \label{claim:wt_dist_for_size_of_vanishing_ideal}
\begin{align*}
        \sum_{P\in \RM(m,r)} \exp_2\left(-\epsilon\cdot \binom{m}{\leq r}\right)\cdot \frac{\binom{(1-\wt(P))2^m}{K}}{\binom{2^m}{K}} \leq 1+o(1).
\end{align*}
\end{claim}

\noindent 
Given \autoref{claim:wt_dist_for_size_of_vanishing_ideal}, we see that 
  \begin{align*}
        \E \left[\frac{|\I_r(Z)|}{\exp_2\left(\binom{m}{\leq r}\right)}\right]&= \exp_2(-K)\cdot \left[ \sum_{P\in \RM(m,r)} \exp_2\left(-\epsilon\cdot \binom{m}{\leq r}\right)\cdot \frac{\binom{(1-\wt(P))2^m}{K}}{\binom{2^m}{K}}  \right] \\
        &\leq \exp_2(-K)\cdot(1+o(1)),
  \end{align*}
which concludes the proof of \autoref{lem:dim_to_size_of_vanishing_ideal}.

\medskip

\noindent
Next, we proceed to prove  \autoref{claim:wt_dist_for_size_of_vanishing_ideal}.
\begin{proof}[Proof of \autoref{claim:wt_dist_for_size_of_vanishing_ideal}]
Let $u=\binom{m}{\leq r}$ in the following. Recall that $K=(1-\epsilon) u$. Also,
    \begin{align*}
            \sum_{P\in \RM(m,r)} \exp_2\left(-\epsilon  u\right)\cdot \frac{\binom{(1-\wt(P))2^m}{K}}{\binom{2^m}{K}} & \leq
            \sum_{P\in \RM(m,r)} \exp_2\left(-\epsilon  u\right)\cdot (1-\wt(P))^{(1-\epsilon) u}.
    \end{align*}
    Set $\delta = \left(\frac{1}{\binom{m}{\leq r}}\right)^2$ and let $\RM_{\delta}(m,r) = \{P \in \RM(m,r) \colon |\wt(P)-1/2|\geq \delta/2 \}$, i.e., the set of polynomials with bias at least $\delta$. 
    
    Now, 
    \[
    \sum_{P\in \RM_\delta(m,r)}\exp_2\left(-\epsilon  u\right)\cdot (1-\wt(P))^{(1-\epsilon) u} 
    \]\[\quad\quad\leq 2\cdot \sum_{P \colon \wt(P)\leq 1/2-\delta/2} \exp_2\left(-\epsilon  u\right)\cdot (1-\wt(P))^{(1-\epsilon) u},
    \]
    since the term corresponding to a  polynomial $P$ of weight greater than $1/2+\delta/2$ can be upper bounded by the term corresponding to the polynomial $1+P$ which has weight at most $1/2-\delta/2.$

    Since the RHS of the above equation involves polynomials whose weights are in the interval $[\sfrac{1}{2^r}, \sfrac{1}{2} - \delta]$, we will split this interval into sub-intervals and analyse the contribution from each.
    \begin{align*}
        \operatorname{Low}_i &:= \insquare{\sfrac{1}{2^{i+1}}, \sfrac{1}{2^i}}&\text{for $i = 2,3,\ldots, r-1$},\\
        \operatorname{Med}_i &:= \insquare{(\sfrac{1}{2} - \sfrac{1}{2^i}), (\sfrac{1}{2} - \sfrac{1}{2^{i+1}})}&\text{for $i = 2,3,\ldots, t = \log\frac{1}{\delta}$}.
    \end{align*}
    Let us use the following quantities to denote the number of polynomials with weights in the above intervals:
    \begin{align*}
        L_i &:= \abs{\setdef{P\in\RM(m,r)}{\wt(P)\in \operatorname{Low}_i}}\\
        M_i &:= \abs{\setdef{P\in\RM(m,r)}{\wt(P)\in \operatorname{Med}_i}}.
    \end{align*}
    Therefore,
    \[
        \sum_{P\in \RM_\delta(m,r)} \exp_2\left(-\epsilon  u\right)\cdot (1-\wt(P))^{(1-\epsilon) u} \leq 2\cdot \sum_{P \colon \wt(P)\leq 1/2-\delta/2} \exp_2\left(-\epsilon  u\right)\cdot (1-\wt(P))^{(1-\epsilon) u}
    \]
    \[
        \quad\quad\leq 2\inparen{\sum_{i=2}^{r-1} L_i \cdot \inparen{1 - \sfrac{1}{2^{i+1}}}^k + \sum_{i=2}^{t} M_i \cdot \inparen{\sfrac{1}{2} + \sfrac{1}{2^{i}}}^k}\cdot\frac{1}{\exp_2\inparen{\binom{m}{\leq r} - K}}.
    \]
     By \autoref{claim:bounding_Ti} proved below using the weight distribution bounds of Sberlo and Shpilka (\autoref{thm:SS-low-wt} and \autoref{thm:SS-high-wt}), we have that 
     \[
\inparen{\sum_{i=2}^{r-1} L_i \cdot \inparen{1 - \sfrac{1}{2^{i+1}}}^k + \sum_{i=2}^{t} M_i \cdot \inparen{\sfrac{1}{2} + \sfrac{1}{2^{i}}}^k}\cdot\frac{1}{\exp_2\inparen{\binom{m}{\leq r} - K}} \leq O(\delta^2).
     \]
    
    Hence,
    \begin{align*}
        \sum_{P\in \RM(m,r)} &\exp_2\left(-\epsilon  u\right)\cdot (1-\wt(P))^{(1-\epsilon) u} \\
        &= \sum_{P\in \RM_\delta(m,r)} \exp_2\left(-\epsilon  u\right)\cdot (1-\wt(P))^{(1-\epsilon) u}\\
        & \quad\quad+ \sum_{P\notin \RM_\delta(m,r)} \exp_2\left(-\epsilon  u\right)\cdot (1-\wt(P))^{(1-\epsilon) u}\\
        &\leq O(\delta^2) + \sum_{P\notin \RM_\delta(m,r)} \exp_2\left(-\epsilon  u\right)\cdot (1/2+\delta/2)^{(1-\epsilon)u}\\
        &\leq O(\delta^2) + (1+\delta)^{(1-\epsilon)u}\\
        &\leq O(\delta^2) + \exp(\delta u)\\
        &\leq 1+O(1/u) = 1 + o(1).
    \end{align*}
\end{proof}

It remains to prove the following technical claim.

   \begin{claim}\label{claim:bounding_Ti}
        \begin{enumerate}
            \item \label{claim:bounding_Ti_a} For all $i = 2,\ldots, r-1$, we have
              \[
                %L_i \cdot \inparen{1 - \frac{1}{2^{i+1}}}^k \leq \delta^3 \cdot \exp_2\inparen{\binom{m}{\leq r} - k}.
                \frac{L_i \cdot \inparen{1 - \sfrac{1}{2^{i+1}}}^k}{\exp_2\inparen{\binom{m}{\leq r} - k}} \leq \delta^3.
            \]
            \item \label{claim:bounding_Ti_b} For all $i = 2,\ldots, t = \log\frac{1}{\delta}$, we have
              \[
                %M_i \cdot \inparen{\frac{1}{2} + \frac{1}{2^{i}}^k}\leq \delta^3 \cdot \exp_2\inparen{\binom{m}{\leq r} - k}.
                \frac{M_i \cdot \inparen{\sfrac{1}{2} + \sfrac{1}{2^{i}}}^k}{\exp_2\inparen{\binom{m}{\leq r} - k}} \leq \delta^3.
            \]
        \end{enumerate}
    \end{claim}

    \begin{proof}[Proof of \autoref{claim:bounding_Ti}\eqref{claim:bounding_Ti_a}]
        Note that $(1-\sfrac{1}{2^{i+1}})^{k}\leq \exp(-\sfrac{k}{2^{i+1}}) \leq \exp_2(-\sfrac{k}{2^{i+1}})$. Hence,
        \begin{align*}
            \frac{L_i \cdot (1-\sfrac{1}{2^{i+1}})^{k} }{\exp_2\inparen{\binom{m}{\leq r} - k}} & \leq \frac{\WtDist_{m,r}(2^{-i}) \cdot \exp_2(-\sfrac{k}{2^{i+1}})}{\exp_2\inparen{\binom{m}{\leq r} - k}}.
        \end{align*}
        Using \autoref{thm:SS-low-wt} to bound $\WtDist_{m,r}(2^{-i})$, we get
        \begin{align*}
            \frac{L_i \cdot \inparen{1 - \sfrac{1}{2^{i+1}}}^k}{\exp_2\inparen{\binom{m}{\leq r} - k}} & \leq \exp_2\left(O(m^4) + O(i\gamma^{i-1})\cdot \binom{m}{\leq r} - \sfrac{k}{2^{i+1}} - \binom{m}{\leq r} + k\right)\\
            &=\exp_2\left(O(m^4) - \binom{m}{\leq r}\cdot (1-O(i\gamma^{i-1})) + k\cdot(1-2^{-i-1})\right),
        \end{align*}
        where $\gamma = \sfrac{r}{m}$. If this $\gamma$ is a small enough absolute constant, we can see that the $O(i\gamma^{i-1})$ term is always at most $2^{-i-1}.$ Using this and continuing the computation above, we get
        \begin{align*}
            \frac{L_i \cdot \inparen{1 - \sfrac{1}{2^{i+1}}}^k}{\exp_2\inparen{\binom{m}{\leq r} - k}} & \leq \exp_2\left(O(m^4) - \left(\binom{m}{\leq r}-k\right)\cdot (1-2^{-i-1})\right)\\
            &\leq \exp_2\left(O(m^4) - \varepsilon \binom{m}{\leq r}\cdot (1-2^{-i-1})\right)\\
            &\leq \exp_2\left(O(m^4) - \frac{\varepsilon}{2}\cdot \binom{m}{\leq r}\right) \leq \exp_2\inparen{-\Omega\inparen{\binom{m}{\leq r}}} \leq \delta^3
        \end{align*}
        where for the second inequality above, we used the fact that $k\leq K = (1-\varepsilon)\cdot \binom{m}{\leq r}$.
    \end{proof}

    \begin{proof}[Proof of \autoref{claim:bounding_Ti}\eqref{claim:bounding_Ti_b}]
        Note that 
        \begin{align*}
            (\sfrac{1}{2}+ \sfrac{1}{2}^{i})^{k} & = 2^{-k} \cdot (1 + \sfrac{1}{2^{i-1}})^k \\
            & \leq 2^{-k} \cdot \exp(\sfrac{k}{2^{i-1}}) = 2^{-k} \cdot \exp_2 (\sfrac{k}{2^{i-1}} \cdot \log_2 e)\\
            & =\exp_2(-k\cdot (1 - 2\log_2 e \cdot 2^{-i})).
        \end{align*}
        Using the fact that $M_i \leq \WtDist_{m,r}\inparen{\sfrac{1}{2} - \sfrac{1}{2^{i+1}} }$, we get
        \begin{align*}
            \frac{\WtDist_{m,r}\inparen{\sfrac{1}{2} - \sfrac{1}{2^{i+1}} } \cdot \exp_2(-k\cdot (1-2\log_2 e \cdot 2^{-i})) }{\exp_2\left( \binom{m}{\leq r} - k\right)}\\
            \leq \frac{\exp_2\left(O(m^4) + (1-2^{-O(\max\{\gamma^2 i, \gamma\})})\cdot \binom{m}{\leq r} -k\cdot (1-2\log_2 e \cdot 2^{-i}))\right)}{\exp_2\left( \binom{m}{\leq r} - k\right)}\\
            = \exp_2\left(O(m^4) -\frac{1}{2^{O(\max\{\gamma^2 i, \gamma\})}}\cdot \binom{m}{\leq r} + k\cdot \frac{1}{2^{-(i-1-\log_2\log_2 e)}}\right)
        \end{align*}
        where $\gamma = r/m$, and we used \autoref{thm:SS-high-wt} for the second inequality.

        Note that as long as $\gamma$ is a small enough absolute constant, the $O(\max\{\gamma^2 i, \gamma\})$ term above is at most $i-1-\log_2\log_2 e$ for each $i\in \{2,\ldots,t\}.$ Using this bound, we continue the above computation as follows.
        \begin{align*}
            \frac{M_i \cdot \inparen{\sfrac{1}{2} + \sfrac{1}{2^{i}}}^k}{\exp_2\inparen{\binom{m}{\leq r} - k}} &\leq \exp_2\left(O(m^4) -2^{-O(\max\{\gamma^2 i, \gamma\})}\cdot \left(\binom{m}{\leq r} - k\right)\right)\\
            &\leq \exp_2\left(O(m^4) -2^{-O(\max\{\gamma^2 i, \gamma\})}\cdot \left(\binom{m}{\leq r} - K\right)\right)\\
            &= \exp_2\left(O(m^4) -2^{-O(\max\{\gamma^2 i, \gamma\})}\cdot\varepsilon\cdot \binom{m}{\leq r}\right)\\
            &\leq \exp_2\left(O(m^4) - \Omega\inparen{\sqrt{\binom{m}{\leq r}}}\right) \leq \delta^3,
        \end{align*}
        where for the last inequality we used the fact that $\binom{m}{\leq r} \geq 2^{\Omega(\gamma \log(1/\gamma) m)}$ and hence for small enough absolute constant $\gamma$, we have
        \begin{align*}
        2^{-O(\max\{\gamma^2 i, \gamma\})}\cdot \binom{m}{\leq r} &\geq 2^{-O(\gamma^2 t)}\cdot \binom{m}{\leq r} \geq \exp_2(\Omega(\gamma \log(1/\gamma) m) - O(\gamma^2 \log(1/\delta)))\\
        &= \exp_2(\Omega(\gamma \log(1/\gamma) m) - O(\gamma^3 \log(1/\gamma)m)) = \exp_2(\Omega(m)).  \qedhere
        \end{align*}
    \end{proof}
    This completes the proof of \autoref{lem:dim_to_size_of_vanishing_ideal}. 

\section{Resilience of Reed-Muller codes under erasures and errors}\label{sec:shannon}
\label{sec:resilience_to_channels}

For parameters $m,r$, recall that $E(m,r)$ is the $\binom{m}{\leq r} \times 2^m$-matrix where the columns are indexed by elements in $\F_2^m$, and the $z$-th column is the column vector consisting of \emph{evaluations} of all multilinear $m$-variate degree $\leq r$ monomials on $z$. It is well-known that the matrix $E(m,r)$ is the generator matrix of the $\RM(m,r)$ code, and is also the parity-check matrix of the $\RM(m,m-r-1)$ code. 

\autoref{thm:vanishing-space-random-set} can be re-interpreted as a statement about random sub-matrices of $E(m,r)$ in the regime where $r \leq \gamma m$ for a small absolute constant $\gamma > 0$. 

\begin{corollary}[rank of random sub-matrices of $E(m,r)$]
    \label{cor:random-submatrices-Emr-rank}
    Let $\gamma > 0$ be an absolute constant and $m,r$ growing parameters with $r \leq \gamma m$. Then, for any constant $\epsilon > 0$, a random set of $K = (1 - \epsilon) \binom{m}{\leq r}$ columns of $E(m,r)$ are linearly independent with probability $1 - o(1)$. 
\end{corollary}
\begin{proof}
    A set of columns of $E(m,r)$ indexed by $Z = \set{z_1,\ldots, z_K}$ are linearly dependent if and only if $\I_r(Z)$ has dimension strictly larger than $\binom{m}{\leq r} - K$. \autoref{thm:vanishing-space-random-set} asserts that this happens with only $o(1)$ probability. 
\end{proof}

\subsection{Channels under consideration}

The above corollary can be used to talk about the resilience of Reed-Muller codes under the \emph{erasure} and \emph{error} channels. We first define the precise definitions to be able to state the results accurately. 

\begin{definition}[binary erasure channel (BEC)] 
    \label{defn:BEC}
    The \emph{binary erasure channel} with parameter $\alpha$, denoted by $\BEC_\alpha$, is the channel with input alphabet $\set{0,1}$ where the each binary symbol is ``erased'' (replaced by the `?' symbol) independently with probability $\alpha$.
\end{definition}

A closely related model is where we have a fixed number of random erasures instead of each coordinate being erased with a fixed probability. We will refer to this as the $\BEC^*$ model although this isn't a channel in the traditional sense of altering each coordinate independently. 

\begin{definition}[capped binary erasure channel (BEC\textsuperscript{*})] 
    \label{defn:BEC*}
    The \emph{capped binary erasure channel} with parameter $K$, denoted by $\BEC^*_K$, refers to the channel with input alphabet $\set{0,1}^n$ that replaces a random subset of at most $K$ of the coordinates by the `?' symbol.
\end{definition}
The Binary Symmetric Channel deals with \emph{errors} or \emph{corruptions} as opposed to erasures in the Binary Erasure Channel. 

\begin{definition}[binary symmetric channel (BSC)] The \emph{binary symmetric channel} with parameter $\alpha$, is the channel where the each binary symbol is ``flipped'' (that is, $0$ changed to $1$ and vice-versa) independently with probability $\alpha$.
\end{definition}
In similar spirit to \autoref{defn:BEC*}, we define the capped binary symmetric channel. 
\begin{definition}[capped binary symmetric channel (BSC\textsuperscript{*})] 
    \label{defn:BSC*}
    The \emph{capped binary symmetric channel} with parameter $K$, denoted by $\BSC^*_K$, refers to the channel with input alphabet $\set{0,1}^n$ that ``flips'' a random subset of at most $K$ of the coordinates.
\end{definition}

In most cases, resilience with respect to $\BEC_\alpha$ (or $\BSC_\alpha$) is the same as resilience with respect to $\BEC^*_K$ (or $\BSC^*_K$)for $K = \alpha n$, by standard concentration inequalities but this might require some subtlety when dealing with codes of rate very close to zero or close to one. For a concrete example, the code $\RM(m,m-1)$ has rate $R = 1 - \sfrac{1}{2^m}$ and is \emph{not} resilient under $\BEC_\alpha$ for $\alpha = (1-R)/2 = \sfrac{1}{2^{m+1}}$ (with constant probability we may have two coordinates erased, and this is not recoverable), but is vacuously resilient under $\BEC^*_K$ for $K = \alpha 2^m$. To avoid these nuances, we will \emph{only} be dealing with the setting of capped channels although this does not make much difference with most of our range of parameters.

\subsubsection*{Notion of ``capacity achieving'' codes with respect to BEC\textsubscript{$\boldsymbol{\alpha}$}} Shannon's seminar work \cite{Shannon1948} showed that, for any constant rate, the supremum of rates of random codes that is resilient to $\BEC_\alpha$ is exactly $R = 1 - \alpha$. That is, for any $\epsilon > 0$, there are codes of rate $1 - \alpha - \epsilon$ that can decode from $\BEC_\alpha$ with decoding error $1 - o(1)$. However, when the rate is very close to zero or one, the situation becomes more nuanced due to the asymptotics involved. 

Dealing specifically with Reed-Muller codes, let us consider the code $\RM(m,m-r-1)$ with $r \leq \sfrac{m}{2}$. This has rate $R = 1 - \frac{\binom{m}{\leq r}}{2^m}$ and can recover from a maximum of $\binom{m}{\leq r}$ errors. Abbe, Shpilka and Wigderson~\cite{AbbeSW2015} defined the notion of ``\emph{capacity achieving under $\BEC$}'' to mean that the code can decode (with $1 - o(1)$ probability) from $(1 - \epsilon) \binom{m}{\leq r}$ erasures for any constant $\epsilon > 0$. We refer the reader to \cite[Section 1]{AbbeSW2015} and \cite[Section 2.2]{SberloS2020} for a nuanced discussion on this.

\subsection{Reed-Muller codes under \texorpdfstring{BEC\textsuperscript{*}}{BEC*}}

An immediate consequence of the above corollary is that $\RM(m,d)$ with $d \geq m(1-\gamma)$ for a small but absolute constant $\gamma > 0$ \emph{achieves capacity} (as defined above) under the Binary Erasure Channel. 

\RestateHighDegRMBEC*
\begin{proof}
    The code $\RM(m,d)$ can recover from erasures on coordinates indexed by $Z \subseteq \F_2^m$ if and only if the corresponding columns of the parity check matrix are linearly independent. Since the parity-check matrix of $\RM(m,d)$ is $E(m,r)$ for $r = m-d-1$, we have from \autoref{cor:random-submatrices-Emr-rank} that a random set of $K$ columns are linearly independent with probability $1 - o(1)$. 
\end{proof}

\subsection{Reed-Muller codes under \texorpdfstring{BSC\textsuperscript{*}}{BSC*}}

The following theorem of Abbe, Shpilka and Wigderson~\cite{AbbeSW2015} provides a method to derive the resilience of certain Reed-Muller codes under the BSC by using the resilience of appropriate Reed-Muller codes under the BEC. Subsequent work of Saptharishi, Shpilka and Volk also showed that these codes also have efficient decoding procedures to recover from random errors. 

\begin{theorem}[\cite{AbbeSW2015}, \cite{SaptharishiSV2017}]
    For any growing parameters $m,r > 0$, if the code $\RM(m,m-2r-2)$ can recover from a subset $Z \subseteq [2^m]$ of erasures, then the code $\RM(m,m-r)$ can efficiently recover from errors on the subset $Z$. 

    In particular, if the code $\RM(m,m-2r-2)$ can recover $K$ random erasures with probability $1 - o(1)$, then the code $\RM(m,m-r)$ can efficiently decode from $K$ random errors. 
\end{theorem}
Applying the above theorem to \autoref{thm:high-deg-RM-under-BEC} yields the following corollary.

\RestateHighDegRMBSC*
\begin{corollary}
    There exists a constant $\gamma_0>0$ such that for all $\gamma<\gamma_0$ and $\epsilon>0$ the following is true. Let $m,r$ be growing parameters with $r = \gamma m$. Then, the code $\RM(m,m-2r-2)$ can be efficiently decode from $K = (1 - \epsilon)\binom{m}{\leq r}$ random errors.\hfill\qedsymbol
\end{corollary}

It is worth noting that the minimum distance of the code $\RM(m,m-2r-2)$ is merely $2^{2r + 2} \ll \binom{m}{\leq r}$ when $r = \gamma m$ for a small enough $\gamma > 0$. Thus, the above corollary shows that high-degree (or high-rate) Reed-Muller codes are resilient to random errors well beyond their minimum distance, and efficiently so. 

{\small
  \bibliographystyle{prahladhurl}
    %\bibliography{/Users/prahladh/Dropbox/Documents/academic/papers/jrnl-names-abb,/Users/prahladh/Dropbox/Documents/academic/papers/prahladhbib,/Users/prahladh/Dropbox/Documents/academic/papers/crossref}
    \bibliography{BHSS}
}

\end{document}